\documentclass[
preprint,%
secnumarabic,%
 amssymb, amsmath,%
 aip,cha,%
]{revtex4-2}

\usepackage{docs}%
\usepackage{bm}%
\usepackage[colorlinks=true,linkcolor=blue]{hyperref}%
\expandafter\ifx\csname package@font\endcsname\relax\else
 \expandafter\expandafter
 \expandafter\usepackage
 \expandafter\expandafter
 \expandafter{\csname package@font\endcsname}%
\fi
\usepackage{graphicx}
\hypersetup{citecolor=blue}

\begin{document}
\title{Topological insulator nanoribbon Josephson junctions: evidence for size effect in transport properties}

\author{Gunta Kunakova}
\affiliation{Quantum Device Physics Laboratory, Department of Microtechnology and Nanoscience, Chalmers University of Technology, SE-41296 G\"oteborg, Sweden}
\affiliation{Institute of Chemical Physics, University of Latvia, Raina Blvd. 19, LV-1586, Riga, Latvia}

\author{Ananthu P. Surendran}
\affiliation{Quantum Device Physics Laboratory, Department of Microtechnology and Nanoscience, Chalmers University of Technology, SE-41296 G\"oteborg, Sweden}

\author{Domenico Montemurro}
\affiliation{Quantum Device Physics Laboratory, Department of Microtechnology and Nanoscience, Chalmers University of Technology, SE-41296 G\"oteborg, Sweden}
\affiliation{Dipartimento  di  Fisica ``Ettore Pancini",  Università  degli  Studi  di  Napoli  Federico  II,  I-80125 Napoli, Italy}

\author{Matteo Salvato}
\affiliation{Dipartimento di Fisica, Università di Roma ``Tor Vergata", 00133 Roma, Italy}

\author{Dmitry Golubev}
\affiliation{QTF Centre of Excellence, Department of Applied Physics,Aalto University, P.O. Box 15100, FI-00076 Aalto, Finland }

\author{Jana Andzane}
\affiliation{Institute of Chemical Physics, University of Latvia, Raina Blvd. 19, LV-1586, Riga, Latvia}

\author{Donats Erts}
\affiliation{Institute of Chemical Physics, University of Latvia, Raina Blvd. 19, LV-1586, Riga, Latvia}

\author{Thilo Bauch}
\affiliation{Quantum Device Physics Laboratory, Department of Microtechnology and Nanoscience, Chalmers University of Technology, SE-41296 G\"oteborg, Sweden}

\author{Floriana Lombardi}
\email{floriana.lombardi@chalmers.se}
\affiliation{Quantum Device Physics Laboratory, Department of Microtechnology and Nanoscience, Chalmers University of Technology, SE-41296 G\"oteborg, Sweden}

%\date{\today}
%\revised{x 2020}

\begin{abstract}
\begin{description}
\item[Abstract] We have used Bi$_2$Se$_3$ nanoribbons, grown by catalyst-free Physical Vapor Deposition to fabricate high quality Josephson junctions with Al superconducting electrodes. In our devices we observe a pronounced reduction of the Josephson critical current density $J_c$ by reducing the width of the junction, which in our case corresponds to the width of the nanoribbon. Because the topological surface states extend over the entire circumference of the nanoribbon, the superconducting transport associated to them is carried by modes on both the top and bottom surfaces of the nanoribbon. We show that the $J_c$ reduction as a function of the nanoribbons width can be accounted for by assuming that only the modes travelling on the top surface contribute to the Josephson transport as we derive by geometrical consideration. This finding is of a great relevance for topological quantum circuitry schemes, since it indicates that the Josephson current is mainly carried by the topological surface states. 
\end{description}
\end{abstract}
\maketitle
\section{Introduction}
\indent The study of the proximity effect between a superconductor and a semiconductor or an unconventional metal, has lately received a dramatic boost due to the increasing possibilities to manufacture a larger variety of interfaces and materials. Novel phenomenology of the proximity effect is currently coming from the integration of semiconducting nanowires, with strong spin orbit coupling, as barriers, as well as the edge and surface states of two-dimensional (2D) and three-dimensional (3D) Topological Insulators (TIs) \cite{Hajer2019, Jauregui2018, Charpentier2017, Galletti2014,Stolyarov2020}, and Dirac semimetals \cite{Li2018b, Li2018a}. In these cases, the Josephson transport properties of the hybrid devices would manifest neat fingerprints related to the formation of Majorana bound states, of great interest for topological quantum computation \cite{Fu2008, Nayak2008, Huang2017}. Lately Superconductor-TI-Superconductor Josephson junctions with 2D \cite{Wiedenmann2016} and 3D TIs have shown a 4$\pi$ periodic Josephson current phase relations \cite{Rokhinson2012, Calvez2018,Schuffelgen2019} which could be associated to the presence of Majorana modes \cite{Dominguez2012} and a gate-tunable Josephson effects \cite{Stehno2019, Ghatak2018a, Jauregui2018, Kayyalha2017a, Cho2013} when the TI is tuned through the TI's Dirac point. Still several aspects of the physics of the Josephson effect related to the topological protected edge/surface states and the contribution of the unavoidable bulk remain to be clarified. In this respect the use of 3D TI nanoribbons could be advantageous because of the reduced number of transport channels involved in the transport. Here the transport is ruled by the quantization of the nanoribbon’s propagation modes, which could give new hints about the Josephson phenomenology associated to the Topological Surface States (TSS).\\
\indent In this work, we have fabricated Josephson junctions by using Bi$_2$Se$_3$ Topological Insulator Nanoribbons (TINR) with a widths spanning from 50~nm to almost a micron. Because the TSS extend over the entire circumference of the TINR, the superconducting transport associated to them is carried by modes on both the top  and bottom surface of the nanoribbon. As shown in Fig.~\ref{DeviceSchemSEM}a, in our TINR Josephson junction the current flows between the superconducting contacts fabricated on the top surface of the nanoribbon. For the TINR with a circumference $C$~=~2$W$~+~2$t$ (where $W$ is the width and $t$ the thickness of the nanoribbon) the transverse momentum $k_y$, perpendicular to the current (see Fig.~\ref{DeviceSchemSEM}), is quantized as:
\begin{equation}
k_y=2\pi(n + 1/2)/C,
\end{equation}
where $n$ is an integer \cite{Jauregui2016}. Therefore, the modes with $k_y$~$\sim$~0 remain on the top surface while the modes with $k_y$~$\gg$~0 are winding around the perimeter of the TINR (see Fig.~\ref{DeviceSchemSEM}a). Here we show that the value of the Josephson current density strongly depends on the junction width, that in our case corresponds to the width of the nanoribbons (Fig.~\ref{DeviceSchemSEM}a). We discuss the possible origin of this phenomenology also in connection with the fact that only small $k_y$ value modes are involved in the Josephson transport, that are the ones which travel on the top surface of the nanoribbon. This number reduces by reducing the  nanoribbon width (see Eq.~1). This finding is of a great relevance since it indicates that a) the Josephson  current is mainly carried by the surface states and b) because of the selectivity in $k_y$, the number of modes involved in the transport scales much faster compared to the width. This is of great relevance for topological quantum circuitry schemes. 
\section{Experimental details and results}
\begin{figure*}
    \centering
    \includegraphics[width=1\linewidth]{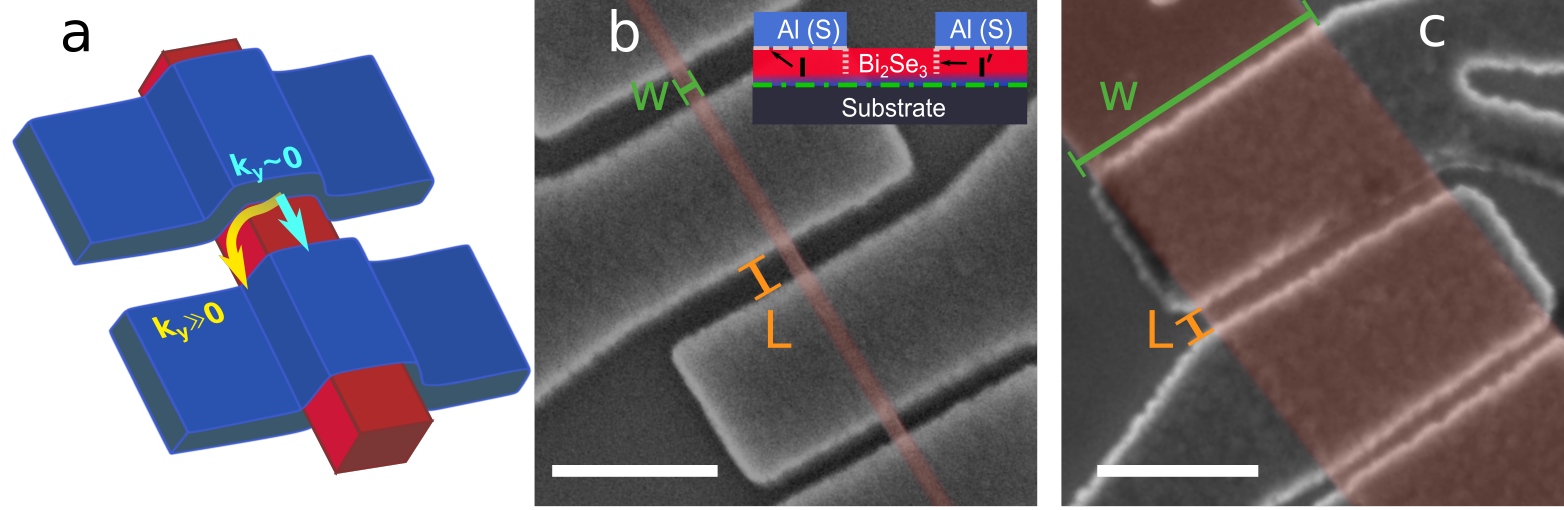}
    \caption{(a) Schematics of a Bi$_2$Se$_3$ nanoribbon Josephson junction. Arrows indicates transport modes carrying supercurrent by the topological surface states at the nanoribbon top surface ($k_y\sim$~0) and around the perimeter ($k_y\gg$~0). (b) and (c) Partly coloured SEM images of the fabricated Josephson junctions of Bi$_2$Se$_3$ nanoribbons with different widths $W$, $L$ indicate length of a junction. Scale bar is 500~nm. Inset in panel (b) is a schematic cross-section of a junction, dot-dashed green line highlights the location of a trivial 2DEG \cite{Kunakova2018}.}
    \label{DeviceSchemSEM}
\end{figure*}

\indent Nanoribbons of Bi$_2$Se$_3$ with a variation of widths from $\sim$50~nm to about 1~$\mu$m were grown as reported in ref.\cite{Andzane2015a}. For the fabrication of Josephson junctions, nanoribbons were transferred to pre-patterned substrates of Si/300~nm SiO$_2$ and SrTiO$_3$ (STO). We used two different TINR growth batches for the two substrates, respectively. For the two batches the carrier densities of the topological surface states measured through Shubnikov de Haas oscillations varied by a factor of 2 \cite{Andzane2015a}. A standard process of electron beam lithography was used, followed by the evaporation of 3 and 80~nm thick layers of Pt and Al respectively. Before the evaporation of the metals, the samples were etched for 30~s by Ar ion milling to remove the native oxide layer of the nanoribbons. SEM images of the fabricated Al/Bi$_2$Se$_3$/Al Josephson junctions for different nanoribbon widths are shown in Fig.~\ref{DeviceSchemSEM}b,c. The Pt interlayer between the Bi$_2$Se$_3$ and Al has an important role in forming of transparent interface, as it was shown in our previous works \cite{Kunakova2019b, Galletti2017, Charpentier2017}.\\Junctions were measured at a base temperature of 19~mK, in an $rf$-filtered dilution refrigerator.\\
\indent Planar Josephson junction can be described as SIS$^\prime$I$^\prime$-N-I$^\prime$S$^\prime$IS. Here S$^\prime$ is the proximized TINR which lies underneath the superconductor (S) - Al, and N represents the not-covered TINR part between the two Al electrodes (see schematics in the inset of Fig.~\ref{DeviceSchemSEM}b). I is the interface of the barrier between the Al and the Bi$_2$Se$_3$ nanoribbon, while I$^\prime$ represents the barrier interface formed between the Bi$_2$Se$_3$ under the Al and that Bi$_2$Se$_3$ of the normal metal region N (nanogap). We have previously demonstrated that both I and I$^\prime$ are highly transparent \cite{Kunakova2019b, Charpentier2017} and that we have full control of  the strength and phenomenology of the proximity effect. When the voltage drop across the Al/Bi$_2$Se$_3$ interface is negligible, compared to that across the TI in the nanogap, as in our case \cite{Aminov1996}, the physics of the planar junction is effectively that of a S$^\prime$I$^\prime$-N-I$^\prime$S$^\prime$.\\
\begin{figure*}
    \centering
    \includegraphics[width=0.9\linewidth]{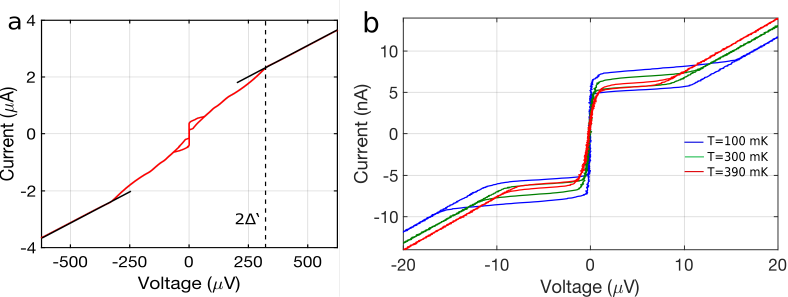}
    \caption{(a) Current – voltage characteristic of a Bi$_2$Se$_3$ nanoribbon junction ($I_c$~=~0.36~$\mu$A, $L$~=~70~nm, $t$~=~16~nm, $W$~=~430~nm) measured at $T=20$~mK, the solid black lines are linear fits of the IVC at high bias voltages. The departure from linearity observed at V~=~340~$\mu$V corresponds to twice the induced gap, with gap $\Delta^\prime$~=~170~$\mu$V (see the dashed line). (b) Low bias current voltage characteristics at various temperatures $T = 100, 300, 390$~mK of a Bi$_2$Se$_3$ nanoribbon junction ($I_c$~=~10~nA, $L$~=~70~nm, $t$~=~13~nm, $W$~=~60~nm). The hysteretic current voltage characteristics develop a finite resistance in the superconducting branch for increasing temperature.
   % and (c) $J_c$ and $R_N(W t)$ as a function of the nanoribbon width $W$ for various junction lengths, respectively.
   }
    \label{Plots_IV}
\end{figure*}
\indent A typical current-voltage characteristic (IVC) of one of the TINR based Josephson junctions is shown in Fig.~\ref{Plots_IV}a. The IVCs of our junctions have a hysteresis which points to an increased electron temperature as the junction is switched to the resistive state \cite{Courtois2008}. A clear gap structure is seen at 2$\Delta^\prime$ (where $\Delta^\prime$ is the induced gap of the order of 170$\mu$V); several bumps in the IVC at 2$\Delta^\prime/n$ are also visible: as we have previously demonstrated they are connected to multiple Andreev reflections that are made possible by the high transparency of the I and I$^\prime$ barriers \cite{Kunakova2019b}. The IVCs measured at various temperatures are shown in Fig. \ref{Plots_IV}b. At higher temperatures we observe a finite resistance in the supercurrent branch. This could be attributed to premature switching and consecutive retrapping of the superconducting phase difference \cite{Krasnov2005,Kivioja2005,Massarotti2015,Longobardi2012}. The so-called phase diffusion regime is characteristic for moderate to low quality factor Josephson junction. The quality factor can be estimated using $Q=\omega_P RC$, with the plasma frequency $\omega_P = \sqrt{2\pi I_c/\Phi_0 C}$, where $\Phi_0=h/2e$ is the superconductive flux quantum, and $R\simeq100~\Omega$ is the real part of the shunting admittance of a dc biased Josephson junction \cite{Bauch2005,Bauch57}. Here the capacitance is dominated by the shunting capacitance through the substrate.\\
\begin{figure}
    \centering
    \includegraphics[width=0.65\linewidth]{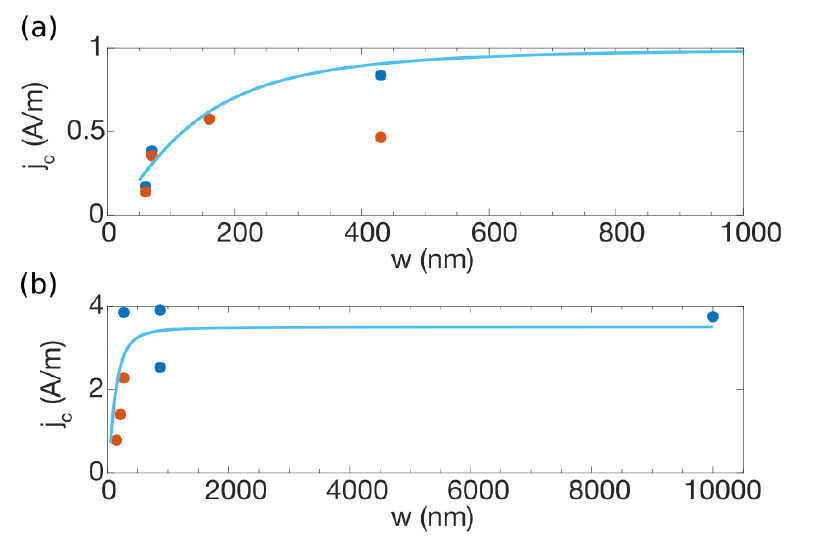}
    \caption{(a) Critical current density as a function of TINR width for Josephson junction devices realized from the same growth batch on a Si/SiO$_2$ substrate. (b) Critical current density as a function of TINR width for Josephson junctions realized from a second growth batch (different from those shown in panel (a)) on a SrTiO$_3$ substrate. All measurements were performed at $T=20$~mK. The blue and red dots are for junction lengths $50-80$~nm and $100-110$~nm, respectively.}
    \label{jc_w}
\end{figure}
For STO with a relative dielectric contant of 25000 at low temperatures we approximate $C\simeq 1~$pF resulting in a quality factor smaller than one for junctions having a critical current lower than 50~nA. Instead for the Si/SiO$_2$ substrate it is more difficult to estimate the shunting capacitance due to the conducting substrate. However we expect a much smaller capacitance value resulting in quality factors smaller than one for critical currents already below 500 nA.\\
\begin{figure}
    \centering
    \includegraphics[width=0.65\linewidth]{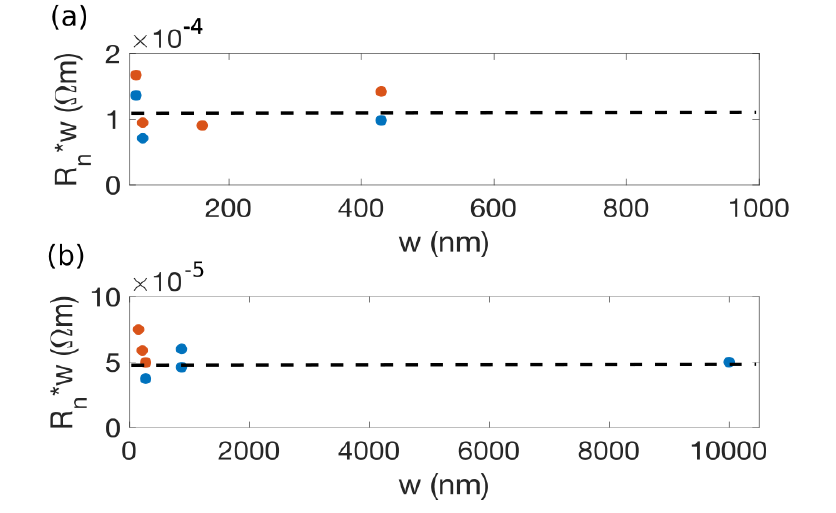}
    \caption{(a) Specific resistance as a function of TINR width for Josephson junction devices realized from the same growth batch on a Si/SiO$_2$ substrate. (b) Specific resistance as a function of TINR width for Josephson junctions realized from a second growth batch (different from those shown in panel (a)) on a SrTiO$_3$ substrate. All measurements were performed at $T=20$~mK. The blue and red dots are for junction lengths $50-80$~nm and $100-110$~nm, respectively.}
    \label{Rn_w}
\end{figure}
The value of the critical current of the junction $I_c$ is obtained from the forward scan, and the critical current density $J_c$ is calculated accordingly by dividing $I_c$ by the width of the nanoribbon $(W)$. The normal state resistance $R_N$ is determined by the inverse of the slope, calculated from the IVC region at voltages above 2$\Delta^\prime$ in S$^\prime$ (represented by the area of Bi$_2$Se$_3$ underneath the Al). Fig.~\ref{jc_w}a,b show the dependence of the $J_c$ as a function of the $W$ for devices with different lengths $L$ fabricated on Si/SiO$_2$ and STO substrates, respectively.  In Fig. \ref{jc_w}b we have included a $10~\mu$m wide junction where the Al electrodes are pattered along the length of TINR; for this device no winding modes are expected to contribute to the Josephson transport. One clearly sees that if the length is fixed the $J_c$ sharply decreases as a function of the width $W$ of the nanoribbon. We note that the value of $J_c$ can change by a factor of 5-6 by going from the narrowest nanoribbons of 60~nm to the widest ones. In contrast, 	as shown in Fig.~\ref{Rn_w}a,b, the specific resistance, obtained by considering the product $R_N\times(W)$ for the devices on the Si/SiO$_2$ and STO substrate, respectively, is almost independent of $W$. This fact allows to exclude that the reduction of the $J_c$ at small widths $W$ has its origin in strong modifications/deterioration of the junction specific resistance for narrow TINR.\\
\section{Discussion and conclusions}
\begin{figure}
    \centering
    \includegraphics[width=0.65\linewidth]{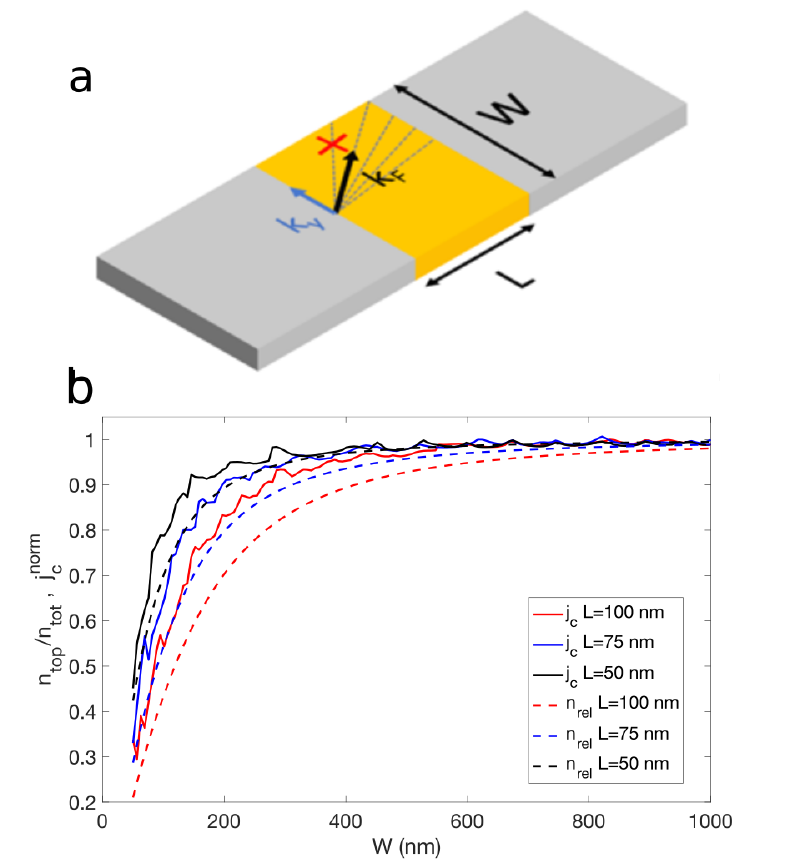}
    \caption{(a) Sketch of a planar TI junction with electrodes separation $L$ and width $W$. The dashed lines indicate quasi-particle trajectories. The maximum transversal momentum for which the transport mode still propagates only on the top surface is indicated by $k_y$. For larger transversal momentum the quasiparticle trajectory has to wind around the TINR and does not contribute to the critical current (red cross). (b) Relative number of transport modes $n_{rel}=n_{top}/n_{tot}$ propagating only on the top surface (dashed lines) and corresponding normalized critical current density (solid lines) as a function of junction width for three different junction lengths.} 
    \label{PlanarJunction}
\end{figure}

\indent What is the origin of this peculiar $J_c$($W$) phenomenology? As we have discussed earlier, the explanation of the phenomenon can have its grounds in the quantization of the nanoribon’s propagation modes.\\
One can derive that the relative number of modes n$_{top}$/n$_{tot}$ travelling only on the top surface reduces with the junction width for a fixed junction length $L$. Here $n_{tot}$ is the total number of modes:
 \begin{equation}
n_{tot}=k_FC/2\pi - 1/2,
\end{equation}
and $n_{top}$ is the number of modes travelling only on the top surface of the nanoribbon:
 \begin{equation}
n_{top}=k_FCW / 4\pi L ((W/2L)^2 + 1)^{-1/2}-1/2.
\end{equation}
The relations above can be obtained from geometric considerations, see Fig.~\ref{PlanarJunction}a. Here $k_F$ is the Fermi vector.\\
In Fig.~\ref{PlanarJunction}b we show the width dependence of the relative number of modes $n_{top}$/$n_{tot}$ for three different junction lengths. The curves of Fig.~\ref{PlanarJunction}b have been obtained by considering $k_F$~=~0.55~nm$^{-1}$, a value which is typical for our nanoribbons \cite{Kunakova2018}. For $L$~=~100~nm we obtain a reduction of the relative number of modes travelling only on the top surface by a factor of 5 when reducing the junction width going from 900~nm to 50~nm.\\
For comparison in Fig. \ref{PlanarJunction} we also show the full calculation of the supercurrent density following ref. \cite{Li2018a} taking into consideration also the angle dependence of the transmission coefficients of each transport mode traveling only on the top surface. We see that the full calculation of $J_C$ and the relative number of transport modes using equation 2 and 3 give the same qualitative behaviour.\\
This dependency qualitatively reproduces the measured $J_c$ $vs$ width dependence shown in Fig.~\ref{jc_w} a,b. Indeed, the solid lines represent the relative number of transport modes for $L=100$~nm. This suggests that only the modes travelling on the top surface contribute to the Josephson current. We note that critical current density of the 10~$\mu$m wide device is in agreement with the saturation value of the expected current density, where the contribution of winding modes to the total Josephson current is negligible.\\
\indent A possible explanation for this finding could be related to the lower mobility of the Dirac states at the interface between the TINR and the substrate. Indeed, our magnetotransport measurements have shown the formation of a trivial 2D gas at the interface with the substrate overlapping with the Dirac states at the nanoribbon bottom (see inset Fig.~\ref{DeviceSchemSEM}b) \cite{Kunakova2018}. This interaction, which leads to a lower mobility (diffusive transport regime) of the Dirac states, might be responsible for the transport modes winding around the nanoribbon (high $k_y$ modes) contributing less to the Josephson transport. Although we can not exclude that the transport through the trivial 2D gas contributes to the Josephson current it would only cause a constant offset of the $J_C$ values without affecting the overall width dependence. We observe that the overall values of $J_C$ for the devices on the STO substrate are higher than those on Si/SiO$_2$. This can be partially attributed to the larger values of the top surface Dirac carrier density of the batch used to realize the devices on the STO substrate. The larger value of the trivial 2D carrier density  we typically observe in devices fabricated on STO substrates \cite{tobepublishedSTO} could be further responsible for the difference observed in the critical current densities between devices on STO and Si/SiO$_2$ substrates.\\
%However there can be other factors contribution to a further. However, there could be other factors contributing to a further reduction of the $J_c$ when  the Josephson energy becomes comparable with the charging energy.
%for lower widths, which can become comparable to the charging energy. This leads to strong fluctuations of the Josephson phase resulting in premature switching of Josephson current at values much smaller than the thermodynamic critical current value \cite{Joyez1999}.\\
\indent To conclude we have fabricated high transparency 3D TINR Josephson junctions showing a peculiar phenomenology that can be associated to the transport through the topological surface states. This is a step forward towards the study of topological superconductivity in few modes devices instrumental for topological quantum computation.
\section*{Acknowledgements}
This work has been supported by the European Union's Horizon 2020 research and innovation programme (grant agreement No. 766714/HiTIMe) and by the European Union's project NANOCOHYBRI  (Cost Action CA 16218). G.K. acknowledges European Regional Development Fund project No 1.1.1.2/VIAA/1/16/198. This work was supported by the European Union H2020 under the Marie Curie Actions (766025-QuESTech). 
\section*{Data availability}
The data that support the findings of this study are available from the corresponding  author upon reasonable request.

\end{document}